\documentclass{ws-p8-50x6-00}

\newcommand{\Tr}{{\rm Tr}}

\newcommand{\ba}{\begin{array}{l}}
\newcommand{\ea}{\end{array}}

\newcommand{\bb}{\begin{thebibliography}}
\newcommand{\eb}{\end{thebibliography}}
\newcommand{\ci}[1]{\cite{#1}}
\newcommand{\lab}[1]{\label{#1}}
\newcommand{\re}[1]{(\ref{#1})}
\newcommand{\Ds}{\displaystyle}

\begin{document}

%%%%% title %%%%%%

 \title{ Light Quarks Beyond Chiral Limit}
\author{M. Musakhanov}
\address{Abdus Salam ICTP, Strada Costiera 11, 34014 Trieste, Italy\\
email: yousufmm@ictp.trieste.it\\
\&
Theoretical  Physics Dept, Tashkent State University,
 Tashkent 700095, Uzbekistan \\
e-mail: yousuf@univer.tashkent.su, yousuf@iaph.silk.org}
\maketitle
\abstracts
{In this talk we discuss an improvement of the 
Diakonov-Petrov QCD Effective Action. We propose  the 
Improved Effective Action, which is derived on the
basis of the Lee-Bardeen results for the quark determinant in the instanton
field. The  Improved Effective Action provides proper
account of the current quark masses, which is particularly important for
strange quarks. This Action is successfully tested by the calculations of the
quark condensate, the masses of the pseudoscalar meson octet
and by axial-anomaly low-energy theorems.}

\section{Introduction}

The properties of the hadrons and their interactions are strongly correlated
with the properties of QCD vacuum.
 All of these properties are concentrated in the Effective Action in terms
of quasiparticles. A very successful
attempt to construct this one was made by Diakonov and Petrov (DP) in 1986
(see recent papers \ci{DPW} and recent detailed
review \ci{SS98} and references therein). Starting from the instanton
model of QCD vacuum, they postulated the Effective Action on the basis of the
interpolation formula for the  well known expression for the light quark
propagator $S_{\pm}$ in the field of the single instanton(anti-instanton)
$
S_{\pm}\,\approx\, S_{0} \, + \, \Phi_{\pm ,
0} \Phi^{\dagger}_{\pm , 0} /im .
$
Here  $S_{0}=(i\hat\partial)^{-1}$ and $\Phi_{\pm , 0} $ are the quark
zero-modes generated by instantons\footnote{$\Phi_{\pm , \lambda} $ is the
eigen-solution of the Dirac equation
$(i\hat\partial + g\hat A_{\pm })\Phi_{\pm , \lambda}\,
=\,\lambda\Phi_{\pm , \lambda}$
in the instanton(anti-instanton) field $A_{\pm , \mu}(x; \xi_{\pm})$.}.
This approach was extended to $N_f =3$ case  \ci{NWZ89a} with account of
the fluctuations of the number of instantons \ci{NWZ89b-KPZ96}.
  
Without any doubts instantons are a very important component of the
QCD vacuum. Their properties are described by the average instanton size
$\rho$ and inter-instanton distance $R$. In 1982 Shuryak
\ci{Shu82} fixed them
phenomenologically as
\be
\rho \,=\, 1/3 \, fm,\, \, \, \, R\,=\, 1 \, fm.
\lab{rhoR}
\ee
From that time the validity of such parameters was confirmed by theoretical
variational calculations
\ci{DP84} and recent lattice simulations of the QCD vacuum (see
recent review \ci{latt}).

The presence of instantons in QCD vacuum very strongly affects light quark
properties, owing to instanton-quark rescattering and consequent generation of
quark-quark interactions.

These effects lead to the formation of the massive constituent
interacting quarks. This implies spontaneous breaking of chiral symmetry
(SBCS), which leads to the collective massless excitations of the QCD
vaccum--pions. The most important degrees of freedom in low-energy QCD  are
these quasiparticles.  So instantons play a leading role in the formation of
the lightest hadrons and their interactions, while the confinment forces are
rather unimportant, probably.

These features of the vacuum are concentrated in the fermionic
determinant ${\det}_N$ (in the field of $N_+$ instantons and  $N_-$ 
antiinstantons), 
which was calculated by
Lee\&Bardeen \ci{LB79} (LB), who
found an amusing result for this quantity:
\be
{\det}_N=\det B, \,\, B_{ij}=
im\delta_{ij} + a_{ji}, \lab{det_N}
\ee
and $a_{ij}$ is the overlapping matrix element of the quark zero-modes
$\Phi_{\pm , 0} $ generated by instantons(antiinstantons).
This matrix element is nonzero only between instantons and antiinstantons
(and vice versa) due to specific chiral properties of the zero-modes
and equal to
\be
a_{-+}=-<\Phi_{- , 0} | i\hat\partial |\Phi_{+ , 0} > .
\lab{a}
\ee
The overlapping of the quark zero-modes provides the propagating of the quarks
by jumping from one instanton to another  one.
So, the determinant of the infinite matrix
was reduced to the
determinant of the finite matrix in the space of \underline{only zero-modes}.
 From Eqs. \re{a}, \re{det_N}
it is clear that for  $N_{+}\neq N_{-}$
$
{\det}_N \sim m^{|N_{+}-N_{-}|}
$
which will strongly suppress the fluctuations of  $|N_{+}-N_{-}|.$
Therefore in final formulas we will assume $ N_{+}=N_{-}=N/2 .$

In \re{det_N}  we observe the competition
between current mass $m$ and overlapping matrix element $a \sim
\rho^{2}R^{-3}$.  With typical instanton sizes $\rho \sim 1/3 fm$ and
inter-instanton distances $R\sim 1fm$, $a$ is of the order of
the strange current quark mass, $m_s = 150 \, MeV$. So in this case it is very
important to take properly into account the current quark mass.

In our previous work \ci{MK,SM,AMT} we showed that the constituent quarks
appear as effective degrees of freedom in the fermionic representation of
$\det B.$ This approach leads to the DP Effective Action with a specific
choice of these degrees of freedom \ci{M99}. 
DP Effective Action is a good tool in the chiral limit but failed 
beyond this limit. This one was checked by the calculations of
the axial-anomaly low energy theorems  \ci{MK,SM,M99}. 
So this Action is hardly applicable to the
strange quarks.

The fermionisation of $\det B$ is not unique procedure and another
fermionic representation of $\det B$  leads to a different choice
of the degrees of freedom in  the Effective Action.
 
Within this approach we are able 
to find an Improved Effective Action which is properly taken into
account current quark masses. 
This Improved Effective Action is checked against direct
calculations of the quark condensate,
the masses of the pseudoscalar meson octet and
against axial-anomaly low energy theorems beyond the chiral limit.

\section{ The Derivation of the  Improved Effective Action  }

The effective action follows from the fermionic representation of the
${\det}_N$  \ci{M99,AMT}. This is not a unique operation. The problem is to
take a proper representation which will define the main  degrees of
freedom in low-energy QCD--constituent quarks.

Let us rewrite the ${\det}_N$ following the idea suggested
in \ci{Tokarev}.  First, by introducing the Grassmanian $(N_{+},N_{-})$ vector
$ \Omega=(u_{1}...u_{N_{+}}, v_{1}...v_{N_{-}}) $ and $ \bar\Omega=(\bar
u_{1}...\bar u_{N_{+}}, \bar v_{1}...\bar v_{N_{-}}) $ we can rewrite 
\bea
{\det}_N &=& \int d\Omega
d\bar\Omega \exp (\bar\Omega B \Omega ) ,\\
\nonumber
\bar\Omega B \Omega  &=& i\sum_{+}m\bar u_{+}u_{+} + i \sum_{-}m\bar
v_{-}v_{-} + \sum_{+-} (\bar u_{+}v_{-}a_{-+} + \bar v_{-}u_{+}a_{+-}) .
\eea

The next step is to introduce   $N_{+},N_{-}$ sources
$\eta=(\eta_{+}, \eta_{-})$
and $N_{-},N_{+}$ sources $\bar\eta=(\bar\eta_{-}, \bar\eta_{+})$
defined as:
$
\bar\eta_{-}=\Phi^{+}_{- , 0} v_{-}(i\overleftarrow{\hat\partial} \,+\,im),
$
\\
$
\bar\eta_{+}=\Phi^{+}_{+ , 0} u_{+}(i\overleftarrow{\hat\partial} \,+\,im),
$
$
\eta_{+}=(i\hat\partial +im) \Phi_{+ , 0} \bar u_{+},
$
$
\eta_{-}=(i\hat\partial +im) \Phi_{- , 0} \bar v_{-}.
$
Then by using the properties of the zero-modes  $\Phi_{\pm , 0}$ and \re{a}
$(\bar\Omega B \Omega)$ and ${\det}_N$  can be rewritten as
\bea
(\bar\Omega B \Omega)&=&- \bar\eta_{+}(i\hat\partial +im)^{-1}\eta_{+} -
\bar\eta_{-}(i\hat\partial +im)^{-1}\eta_{-} 
\\
\nonumber
&-&
- \bar\eta_{-}(i\hat\partial +im)^{-1}\eta_{+} -
\bar\eta_{+}(i\hat\partial +im)^{-1}\eta_{-}
\\
\nonumber
{\det}_N  &=&
\left(\det(i\hat\partial +im)\right)^{-1} \int d\Omega d\bar\Omega
D\psi D\psi^{\dagger}
\exp\int dx [\psi^{\dagger} (x)
(i\hat\partial\,+\,im)\psi (x)
\nonumber\\
&+& \bar\eta_{+} (x)\psi (x) \,+\,\bar\eta_{-} (x)\psi (x)
\,+\,\psi^{\dagger} (x)\eta_{+} (x)\,+\,\psi^{\dagger} (x)\eta_{-} (x)]
\eea
The integration  over Grassmanian variables $\Omega$ and $\bar\Omega$
(with the account  of the $N_{f}$ flavors
${\det}_N = \prod_{f}\det B_{f}$)
 provides the fermionized representation of Lee\& Bardeen's result
for ${\det}_N$ in the form:
\bea
{\det}_N =
\int D\psi D\psi^{\dagger} 
\prod_{f} \exp(\int d^4 x
\psi_{f}^{\dagger}(i\hat\partial \,+\, im_{f})\psi_{f})
\\\nonumber
\times \prod_{+}^{N_{+}} V_{+}[\psi_{f}^{\dagger} ,\psi_{f}]
\prod_{-}^{N_{-}}V_{-}[\psi_{f}^{\dagger},\psi_{f}] ,
\label{part-func}
\\
V_{\pm}[\psi_{f}^{\dagger} ,\psi_{f}]
= \int d^4 x \left(\psi_{f}^{\dagger} (x) (i\hat\partial\,+\,im_{f})
\Phi_{\pm , 0} (x; \xi_{\pm})\right)
\\\nonumber
\times\int d^4 y
\left(\Phi_{\pm , 0} ^\dagger (y; \xi_{\pm} )
(i\hat\partial \,-\,im_{f})\psi_{f} (y)\right).
\lab{V}
\eea
Eq. \re{part-func} exactly represents the fermionic determinant in terms
of constituent quarks $\psi_{f}$. This expression differs from the ansatz
on the fixed $N$ partition function
postulated by DP by another account of the current mass of quarks.

  Keeping in mind the low density of the instanton media, which
allows independent averaging over positions and orientations
of the instantons, Eq. \re{part-func} leads to the partition function
\be
 Z_N = \int D\psi D\psi^\dagger
 \exp\left(\int d^4 x
\sum_{f}\psi_{f}^{\dagger}(i\hat\partial \,+\, im_{f})\psi_{f}\right)
 \,  W_{+}^{N_+}  \, W_{-}^{N_-},
\lab{Z_NW}
\ee
where at arbitrary $N_f$ the integration over $\xi_{\pm}$ leads to
\be
W_\pm =\int d \xi_{\pm}\prod_{f}V_{\pm}[\psi_{f}^{\dagger} \psi_{f}]
 \,=\,
(-i)^{N_{f}}\left(  \frac{4\pi^2
\rho^2}{N_c} \right)^{N_f} \int \frac{d^4 z}{V}
 {\det}_{f}\left( i J_\pm  (z)\right) .
\lab{W}
\ee
Here
\bea
J_\pm (z)_{fg} &=& \int \frac {d^4 k_{1}d^4 k_{2} }{(2\pi )^8 }
e^{i(k_{2}-k_{1})z)}
F(k_{1}) F(k_{2})
 \, \psi^\dagger_f (k_{1})
(\frac{1\pm\gamma_{5}}{2}
\nonumber\\
&+&\frac{ im_g \hat k_{2}}{k_{2}^{2}}\frac{1\mp\gamma_{5}}{2}
\,-\,\frac{ im_f \hat k_{1}}{k_{1}^2}\frac{1\pm\gamma_{5}}{2}
+ \frac{m_f m_g \hat k_{1}\hat k_{2}}{k_{1}^{2}k_{2}^{2}}
\frac{1\mp\gamma_{5}}{2}) \psi_g (k_{2}).\,\,\,\,\,\,\,\,\,\,
\lab{J_pm}
\eea
The form-factor $F(k)$ is related  to the zero--mode
wave function in momentum space $\Phi_\pm (k; \xi_{\pm}) $ \ci{DPW}.

The two remarkable formulas
\be
 (ab)^N = \int d\lambda \exp (N ln \frac {aN} {\lambda} - N +\lambda b )
\,\,(N >>1).
\lab{ab^N}
\ee
and
\be
\exp (\lambda \det [i A] ) =
\int d\Phi \exp\left[ - (N_f - 1) \lambda^{-\frac{1}{N_f - 1}}
(\det\Phi )^{\frac{1}{N_f - 1}} + i tr (\Phi A) \right]
\lab{expA}
\ee
have been used here.
Formula \re{ab^N} leads to exponentiation, while \re{expA}  leads to
the bosonization of the  partition function \re{Z_NW}.
Starting from these formulas, we find our main result
\be
Z_N
=\int d\lambda_{+} d\lambda_{-}
 D\Phi_{+}D\Phi_{-}
\exp\left(-W[\lambda_{+},\Phi_{+};\lambda_{-},\Phi_{-}]\right),
\label{Z}
\ee
where
\bea
W[\lambda_{+},\Phi_{+};\lambda_{-},\Phi_{-}] = -\sum_{\pm}
 \left( N_{\pm} \ln [\left(  \frac{4\pi^2
\rho^2}{ N_c} \right)^{N_f}\frac{N_{\pm}}{V\lambda_{\pm}}] - N_{\pm}\right)
+ w_{\Phi} + w_{\psi},
\nonumber
 \\
w_{\Phi} = \int d^4 x \sum_{\pm}
 (N_f - 1) \lambda_{\pm}^{-\frac{1}{N_f - 1}}
(\det\Phi_{\pm} )^{\frac{1}{N_f - 1}}   , \,\,\,\,\,\,\,\,\,\,\,\,\,\,\,\,\,\,
\,\,\,\,\,\,\,\,\,\,\,\,\,\,\,\,\,\,\,\,\,\,\,\,\,\,\,\,\,\,\,\,\,\,\,\,\,\,\,
\nonumber
 \\
w_{\psi} = -
\Tr \ln (-\hat k  + im_{f} + i
F(k_{1})F(k_{2}) \sum_{\pm}
\Phi_{\pm , gf}(k_{1}-k_{2})(\frac{1\pm\gamma_{5}}{2} + \,\,\,\,\,\,\,\,\,\,
\nonumber
 \\
+\frac{ im_{g}\hat k_{2}}{k_{2}^{2}}\frac{1\mp\gamma_{5}}{2}
\,-\,\frac{ im_{f}\hat k_{1}}{k_{1}^2}\frac{1\pm\gamma_{5}}{2}
+ \frac{m_{f}m_{g}\hat k_{1}\hat k_{2}}{k_{1}^{2}k_{2}^{2}}
\frac{1\mp\gamma_{5}}{2}) (- \hat k + i m_f )^{-1}). \,\,\,\,\,\,\,\,\,\,
\lab{W1}
\eea
Variation of the total action $W[\lambda_{+},\Phi_+;\lambda_{-}\Phi_-]$  over
$\lambda_{\pm},\,\Phi_{\pm}$ must vanish in the common saddle-point.
In  this point
$\lambda_{\pm}=\lambda ,\,\,\Phi_{\pm ,fg}= \Phi_{\pm
,fg}(0)=M_{f}\delta_{fg} .$
This condition leads to the
momentum dependent constituent mass $M_{f}(k)=M_{f}F^{2}(k)$
are calculated from the saddle-point equations 
\bea
\frac{4VN_{c}}{N}\int \frac{d^4 k}{(2\pi)^{4}}\frac{M_{0}^{2}F^{4}(k)}
{k^{2} + M_{0}^{2}F^{4}(k)}
&=& 1,
\lab{saddle1}\\
2\gamma \int k^2 dk^2 \frac{k^2 F^4 (k)}{(k^2 + M_0^2
F^4 (k))^2}
&=& \int k^2 dk^2
\frac{(M_0^2 F^4 (k)- k^2 ) F^2 (k)}{(k^2 + M_0^2 F^4 (k))^2} .
\lab{gamma} \eea
We keep here only $O(m_{f})$ terms and
define $M_{f}=M_0 + \gamma m_{f}.$ 
The solution of the saddle-point equations \re{saddle1}, \re{gamma}
corresponding \re{rhoR} are $M_0 =340\, MeV$ 
and $\gamma = -1.75.$

Finally, the constituent  quark propagator has a form:
\be
 S\,=\,(i\hat\partial \,+ \,i(m_{f} \,+ \, M_{f}F^2 ))^{-1},
 \lab{propagator}
\ee
where $M_{f}$  are given by \re{saddle1}, \re{gamma}.

\section{ Tests for the Improved Effective Action}

Improved and DP Effective Actions coincides in chiral limit. So,
we may expect essential differences in the results only beyond
this limit.

We will test \re{W1} by calculating the quark condensate, 
the masses of the pseudoscalar meson octet and
axial-anomaly low-energy theorems, which will be reduced to the
calculations of the specific correlators.
\\
\underline{The quark condensate and the pseudoscalar meson masses.}

First, we calculate the quark condensate by using the evident formula
\bea
i<\psi_{f}^{\dagger}\psi_{f} >\,&=&\,
V^{-1}Z_{N}^{-1}\frac{\partial Z_{N}}{\partial m_{f}}
\nonumber\\
&=&-\frac{\delta W}{\delta m_f}\,=\,-
\sum_{\pm}(\frac{\delta w_{\Phi}}{\delta \Phi_{\pm}} +
\frac{\delta w_{\psi}}{\delta \Phi_{\pm}})|_{\Phi}
\frac{\delta \Phi}{\delta m_f}
\nonumber\\ &+& \Tr i[(-\hat k + im_{f} +iF^{2}M_{f})^{-1} - (-\hat k +
im_{f})^{-1}].
\lab{condensate1}\eea
Another way is to calculate it directly
\bea i<\psi_{f}^{\dagger}\psi_{f} >\,&=&\, \Tr i[(-\hat k + im_{f}
+iF^{2}M_{f})^{-1} - (-\hat k + im_{f})^{-1}].
\lab{condensate2} \eea
The first term in \re{condensate1} vanishes at the saddle-point and we have a
perfect equivalence of the two calculations of the condensate, in contrast
with analogous calculations with the DP Action\ci{DPW,SM,M99}.
With the formula \re{condensate2} we get
\be i<\psi_{f}^{\dagger}\psi_{f} >\,=\,
N_c \int  \frac{k^2dk^2}{4 \pi^2} (\frac{m_f +M_f F^{2}(k)}{k^2+(m_f
+M_f F^{2}(k))^2} - \frac{m_f}{k^2 + m_{f}^2}).
\lab{condensate3}
\ee
Simple
numerical calculations by using \re{saddle1}, \re{gamma} leads to the
\bea
i<\psi^{\dagger}\psi >|_{m=0}&=& 0.0171 \, GeV^3 ,\,\,
\frac{<\psi^{\dagger}\psi >|_{m=0.15}}{<\psi^{\dagger}\psi >|_{m=0}} -1
= -0.5 .
\lab{condensate4}
\eea
 It is clear
from \re{condensate4} that the calculations with Improved 
Effective Action lead to
the expected dependence on the current mass.  

Now I will calculate the masses of the pseudoscalar mesons.
The matrices $\Phi_{\pm}$, whose usual decomposition is
$
\Phi_{\pm} = \exp(\pm \frac{i}{2}\phi )M\sigma \exp(\pm \frac{i}{2}\phi ) ,
$
$\phi$ and $\sigma$ being $N_{f} \times N_{f}$
matrices, describes mesons
and $M_{fg}=M_f \delta_{fg}$.
At the saddle-point $\sigma = 1,\, \phi = 0$. The
usual decomposition for the  pseudoscalar fields
$\phi = \sum_{0}^{8}\lambda_{i}\phi_{i}$
may be used.
These mesons are considered as a small fluctuation near the saddle point.
The linear on $\phi$ term in \re{W1}
is equal zero at the saddle point and we
consider the next $O(\phi^2 )$  terms. There is no contribution
from $w_{\Phi}$ since
$w_{\Phi} \sim (\prod_f M_f )^{\frac{1}{N_f - 1}} = const.$
On the other hand  $w_{\psi}$ makes a contributions like one-point
and two-points diagrams
\bea
\ba
w_{\psi}=
\\\Ds
 - \Tr\sum_{f }  [ (-\hat k  + i(m_{f}+M_f F^{2}(k )))^{-1} F^{2}(k)
\frac{(-i)}{8}(M \phi^2 + \phi^2  M + 2 \phi M \phi)_{ff}
\\
+\frac{1}{8} \sum_{g }( (-\hat k_1  + i(m_{f}+M_f F^{2}(k_1 )))^{-1}
F(k_{1})F(k_{2}) (M \phi (p ) + \phi (p ) M)_{fg}
\\\Ds
\times
\gamma_{5}(1 - \frac{ im_{g}\hat k_{2}}{k_{2}^{2}}
- \frac{ im_{f}\hat k_{1}}{k_{1}^2})
(-\hat k_2  + i(m_{g}+M_g F^{2}(k_2 )))^{-1}
\\\Ds
\times
F(k_{2})F(k_{1})
(M \phi (-p ) + \phi (-p ) M)_{gf}
- \frac{ im_{f}\hat k_{1}}{k_{1}^2})]
\ea
\lab{action-phi^2}
\eea
where $p = k_1 - k_2 .$ First $p=0$-term is considered.
From the saddle-point eqs. \re{saddle1}, \re{gamma} we get
\bea
w_{\psi}|_{p=0}
= \frac{1}{2}i<\psi^\dagger \psi > (2m \sum_{1}^{3}\phi_{i}^{2}
+ (m_s +m) \sum_{4}^{7}\phi_{i}^{2} +\frac{2}{3}(2m_s +m)\phi_{8}^{2}).
 \eea
Then
$ m_{K}^{2}/m_{\pi}^{2} = (m_s +m)/{2m}=13.5$
and
$m_{\eta}^{2}/m_{\pi}^{2} = (2m_s +m)/{3m}=17.7$
where $m_u = m_d = m \sim 5 \, MeV$ and $m_s \sim 130\, MeV$ were used.
The  experimental values of the masses lead to
$\frac{m_{K}^{2}}{m_{\pi}^{2}} =13.4$
and
$\frac{m_{\eta}^{2}}{m_{\pi}^{2}} =16.5.$

The calculation of the
$p^2$-term in $w_\psi$ provides a normalization factor.
Since $p=0$-term is in the order of $m$ (and its $O(m^2 )$ contributions
were neglected) we calculate  $p^2$-term in the chiral limit $m=0$.  Then
$p^2$-term in $w_\psi$ is extracted from
\bea
\nonumber
&\,& \frac{1}{2} \phi(p)\phi(-p) 4N_c \int \frac{d^4 k}{(2\pi )^4}
\frac{ M^{2}_{0} F^{2}(k_{2})F^{2}(k_{1})
(k_1 k_2 +M^{2}_0 F^{2}(k_1 )F^{2}(k_2 ))}
{(k_{1}^2+M^{2}_0 F^{4}(k_1 ))(k_{2}^2+M^{2}_0 F^{4}(k_2))}\\
&=& \frac{1}{2} \phi(p)\phi(-p)(\frac{N}{V} - f^{2}_{\pi} p^2)
\eea
Combining this result with the calculations of the $p=0$-term we get
$$ m^{2}_{\pi}= \frac{i<\psi^\dagger \psi > 2 m}{ f^{2}_{\pi}}$$
and all of other masses of octet of the pseudoscalar mesons.
Therefore, Improved Effective Action successfully reproduce the current
algebra results.

The next test is related to axial-anomaly low-energy theorems
($LET$)\ci{Shi}.  These theorems were used in \ci{MK,SM} to check DP Effective
Action. The DP Effective Action was able to reproduce $LET$ only in chiral
limit and failed beyond this limit.
\\ 
\underline {$LET1.$}
Nonvanishing of the $\eta '$ meson mass $m_{\eta '}$
even in chiral
limit (due to axial anomaly)
implies that for real photons the matrix element of
the divergence of the singlet
axial current vanishes in the $q^{2}\,<< \, m_{\eta '}^{2}$ limit,
giving rise to the following low energy theorem  ($LET1$):
\be
 \langle 0|N_f \frac{g^2}{16\pi^2}G\tilde G | 2\gamma \rangle
+ 2i\sum_{f} m_{f}
\langle 0|\psi^{\dagger}_{f}\gamma_{5}\psi_{f}| 2\gamma \rangle =
 N_c \frac{e^2}{4\pi^2} \sum_{f} Q^{2}_{f}
F^{(1)}\tilde F^{(2)},
\lab{theorem}
\ee
where $F_{\mu\nu}^{(i)}= \epsilon_{i,\mu}q_{i,\nu} - \epsilon_{i,\nu}q_{i,\mu}$
and $q_i ,\, \epsilon_{i}\,(i=1,2)$ are the momentum and
polarization vectors of photons and $q = q_{1} + q_{2}$ respectively.
Eq. \re{theorem} is an exact low energy relation, which cannot be fulfilled in
the framework of perturbation theory.
 Only a nonperturbative contribution of
order $g^{-2}$  - as the one provided by instantons - may cancel
the factor $g^2$ at the first term of the l.h.s..
The first term of the l.h.s. in \re{theorem} is calculated from three-point
correlator of the operator $g^2 G \tilde G$ and two operators of the
electromagnetic currents. 
 The operator $g^2 G\tilde G$ generates the vertex $i\,f\,F^{2}\, M_f
\,N_{f}^{-1}\,\gamma_5$,  where $f(q^{2})$  is a momentum representation
of the instanton contribution in the operator $g^2 G\tilde G(x)$ and
$f(0)=32\pi^2$ \ci{MK,SM}.
At small $q^2$ this vertex is reduces to
\be
32\pi^2  i\,\,F^{2}\, M_{f}\, N_{f}^{-1}\,\gamma_5
\lab{vertex}
\ee
Then, the three-angular diagrams corresponding to  the
the anomaly contribution (the first term in the l.h.s.
of \re{theorem}), with vertices \re{vertex},
 $eQ_{f}\gamma_{\mu}$ and propagator
\re{propagator} leads to
\be
2i
N_c  e^2 {Q^{2}_{f}} F^{(1)}\tilde F^{(2)}
\Gamma_{f},
\lab{tau1}
\ee
where $\Gamma_{f}$, the factor coming
from the diagram of the process considered, may be calculated analytically if
we approximate the form factor $F$ by $1.$ In this approximation
\be
\Gamma_{f}\,=\,\frac{M_{f}}{8\pi^2 (M_{f}+m_f)}
\lab{FF}
\ee
In the same approximation the current mass contribution (the second term in
the l.h.s.  of \re{theorem})
leads to
\be
2i
N_c  e^2 {Q^{2}_{f}} F^{(1)}\tilde F^{(2)}
\frac{m_{f}}{8\pi^2 (M_{f}+m_f)}
\lab{tau2}
\ee
At the next step we combine \re{tau1}  and \re{tau2} and sum up over
flavors.
As a result, the l.h.s. and the r.h.s of eq. \re{theorem}
coincide with each other.
So, Improved Effective Action immediately fulfills low energy-theorem $LET1$
\re{theorem} even beyond the chiral limit in contrast with
DP Effective Action result \ci{SM}.

If we take into account the form factor $F$ in \re{tau1}, \re{tau2} and give
the model parameters the values \re{rhoR}, we find\ci{MK} a variation of $\sim
17\%$.
\\
\underline {$LET2$, $LET3.$}
Further
tests for Improved Effective Action can be obtained from the matrix
elements of the divergence of the singlet
axial current  between vacuum and meson states. Neglecting
$O(m^2 )$ terms, we get the following equations:
\begin{eqnarray}
 \langle 0|N_f \frac{g^2}{16\pi^2}G\tilde G | \eta \rangle &=&
 - 2 i m_{s}\langle 0|\psi^{\dagger}_{s}\gamma_{5}\psi_{s} | \eta \rangle, \
\lab{relation1}
\\
\langle 0|N_f \frac{g^2}{16\pi^2}G\tilde G | \pi^{0} \rangle &=&
- i (m_{u}-m_{d})
\langle 0|\psi^{\dagger}\tau_{3}\gamma_{5}\psi | \pi^{0} \rangle,
\lab{relation2}
\end{eqnarray}
which we call $LET2$ and $LET3$ respectively.
These matrix elements are reduced to two-point correlators.
It is rather easy to show that Improved Effective Action satisifies
$LET2$ \re{relation1} and $LET3$ \re{relation2}.

From previous considerations it follows that the factor  $g^{2}G\tilde G$
generates the vertex $i M f F^{2}\gamma_{5}N_{f}^{-1}$ and the $\eta$-meson
gives rise to $i M_s \lambda_{8} F^{2}\gamma_{5}.$ The structure of the mass
matrix $M$ is
\be
M \, = \, M_{0} \, + \, \gamma (m_{s}(\frac{1}{3} -\frac{1}{\surd 3}\lambda_{8})
+ m_{u}\frac{1+\tau_{3}}{2} + m_{d}\frac{1-\tau_{3}}{2}).
\lab{M}
\ee
Then at small $q$ (and neglecting $m_{u,d}$)
\bea
 \langle 0|N_f \frac{g^2}{16\pi^2}G\tilde G | \eta \rangle
&=&
-\frac{16 N_{c}}{\surd 3} \int \frac{d^4 k}{(2\pi )^4}F^4 (k)
[\frac{M^{2}_{s}}{((M_{s}F^2 (k)+ m_{s})^{2}  + k^2)^2}
\nonumber \\ &-&
 \frac{M^{2}_{0}}{(M^{2}_{0}F^4 (k)  + k^2)^2}] .
\lab{me1}
\eea
Expanding \re{me1} over $m_s$, we get
\be
 \langle 0|N_f \frac{g^2}{16\pi^2}G\tilde G | \eta \rangle   =
-\frac{16 N_{c} m_{s}}{\surd 3} \int \frac{d^4 k}{(2\pi )^4}F^4 (k)M_0
\frac{2\gamma k^2  - 2 M_0^2 F^2 (k) }{(k^2 + M_0^2 F^4 (k))^2}.
\lab{me11}
\ee
From eq. \re{gamma} for the $\gamma$-factor we find that
\be
2\gamma \int \frac{d^4 k}{(2\pi )^4}
\frac{k^2 F^4 (k)}{(k^2 + M_0^2 F^4 (k))^2} = \int \frac{d^4 k}{(2\pi
)^4} \frac{(M_0^2 F^4 (k)- k^2 ) F^2 (k)}{(k^2 + M_0^2 F^4 (k))^2}
\lab{approx}
\ee
It is clear now that by using \re{approx} the l.h.s of
\re{relation1} is reduced to the r.h.s. of \re{relation1}, which is equal to
\be
- 2im_{s}\langle 0|\psi^{\dagger}_{s}\gamma_{5}\psi_{s} | \eta \rangle  =
\frac{16N_{c} m_{s} }{\surd 3}  \int   \frac{d^4 k}{(2\pi )^4} \frac{M_0
 F^2 (k)}{k^2 + M_0^2 F^4 (k)}.
\lab{me2}
\ee
The calculations with $LET3$
\re{relation2} are almost the same as with $LET2$ \re{relation1}. Again, by
using \re{approx} l.h.s. and r.h.s. of \re{relation2} coincide with each
other.  Hence Improved Effective Action satisfies $LET2$ and $LET3$,
\re{relation1} and \re{relation2} respectively.  For comparison, DP Effective
Action failed to reproduce these $LET2$ and $LET3$ \ci{SM}.

Therefore, Improved Effective Action generates correct current mass dependence
of the vacuum quark condensate, reproduces current algebra results for the
masses of the pseudoscalar meson octet,
satisfies low-energy theorems $LET2$, $LET3$ for the two-point correlators
\re{relation1} and \re{relation2} respectively and also satisfies $LET1$ for
the three-point correlator \re{theorem} even beyond chiral limit. We conclude
that Improved Effective Action works properly beyond chiral limit and provides
the background for taking into account strange quarks.
\\ \\
 I would like to thank the organizers of the Second International Conference on
 Perspectives in Hadronic Physics for the invitation and M.Polyakov and 
C.Weiss for the useful discussions. 
I acknowledge a partial support by ICTP, the grant
INTAS-96-0597ext and the State Committee for Science
and Technology of Uzbekistan.


\begin{thebibliography}{99}

\bibitem{DPW} D.I. Diakonov,  M.V. Polyakov, C. Weiss,
{\it  Nucl. Phys. } {\bf  B 461}, 539 (1996);
 Dmitri Diakonov, hep-ph/9602375, hep-ph/9802298

\bibitem{SS98} T. Schaefer, E. Shuryak, 
{\it  Rev.Mod.Phys.}{\bf 70},323 (1998)

\bibitem{NWZ89a} M.A.Nowak, J.J.M.Verbaarschot, I.Zahed,
 {\it  Nucl.Phys. }{\bf  B 324},1(1989)

\bibitem{NWZ89b-KPZ96} M.A.Nowak, J.J.M.Verbaarschot, I.Zahed,
{\it  Phys.  Lett.} {\bf  B 228}, 251 (1989);
M.Kacir, M.Prakash, I.Zahed,  hep-ph/9602314
 
\bibitem{Shu82}E. V. Shuryak,
{\it  Nucl. Phys. } {\bf  B 203}, 93, 116 (1982)

\bibitem{DP84} D. Diakonov and V. Petrov,
{\it  Nucl. Phys. } {\bf  B 245}, 259 (1984)

\bibitem{latt}
T. De Grand, A. Hasenfratz, T. Kovacs,
 {\it Progr.Theor. Phys. Suppl.}{\bf 131}, 573 (1998)
 
\bibitem{LB79} C. Lee, W. A. Bardeen,
{\it  Nucl. Phys. } {\bf  B 153}, 210 (1979)

\bibitem{MK} M. M. Musakhanov, F. C. Khanna,
{\it  Phys.  Lett.} {\bf  B 395}, 298 (1997)

\bibitem{SM} E. Di Salvo, M.M. Musakhanov,
 {\it Europ.Phys.J. }{\bf C 5}, 501 (1998)

\bibitem{M99} M.M. Musakhanov, hep-ph/9810295, {\it Europ. Phys. J.}
{\bf C9}, 235 (1999)

\bibitem{AMT} F. Araki, M. Musakhanov, H. Toki, hep-ph/9808290

\bibitem{Tokarev}  V.F. Tokarev,
{\it Soviet J. Teor. Math. Phys.}, {\bf  73}, 223 (1987)

\bibitem{Shi} M.A. Shifman,  {\it Sov. Phys. Usp. } {\bf 32}, 289 (1989)

\bibitem{VZNS83}
A.I.Vainshtein, V.I.Zakharov, V.A.Novikov, M.A.Shifman,
Sov.J.Nucl.Phys. {\bf 39}, 77 (1984).

\end{thebibliography}
\end{document}